\documentclass[journal,final,12pt,journal]{IEEEtran}
\usepackage{xcolor,soul,framed} 
\usepackage{multirow}
\colorlet{shadecolor}{yellow}
\usepackage[pdftex]{graphicx}
\graphicspath{ {./figures/} }
\DeclareGraphicsExtensions{.pdf,.jpeg,.png, .jpg}

\usepackage[cmex10]{amsmath}
\usepackage{array}
\usepackage{mdwmath}
\usepackage{mdwtab}
\usepackage{eqparbox}
\usepackage{url}
\usepackage{cite}

\usepackage{array}
\newcolumntype{P}[1]{>{\centering\arraybackslash}m{#1}}

\definecolor{Gray}{gray}{0.85}
\newcolumntype{a}{>{\columncolor{Gray}}c}

\usepackage{balance}

\begin{document}
\bstctlcite{IEEEexample:BSTcontrol}
\title{Performance Requirements of Advanced \\ Healthcare Services over  Future Cellular Systems}
\author{\small Giulia Cisotto, Edoardo Casarin, and Stefano Tomasin

\thanks{Giulia Cisotto is with University of Padova (Italy) and the National Centre for Neurology and Psychiatry, Tokyo (Japan); Edoardo Casarin and Stefano Tomasin are with University of Padova (Italy).}
}

\maketitle

\begin{abstract}
The fifth generation (5G) of communication systems has ambitious targets of data rate, end-to-end latency, and connection availability, while the deployment of a new flexible network architecture will spur new applications. E-health and mobile health (m-health) solutions will meet the increasing demand of new, sustainable, and more accessible services beneficial to both  practitioners and the rapidly aging population. This paper aims at defining the technical requirements of future cellular networks to support a variety of advanced healthcare services (e.g., smart hospital, telesurgery, connected ambulances, and monitoring). While 5G will be able to satisfy these requirements, it will also pave the way for future e- and m-health in the sixth-generation (6G) cellular networks.
\end{abstract}


%
\IEEEpeerreviewmaketitle

\baselineskip=12pt

\section*{Introduction}
As life expectancy extends, especially in the most developed countries, new age-related healthcare needs arise: indeed, with age, mobility degrades and the risk of severe brain injuries and neurodegenerative diseases increases. Also, other secondary pathologies and disorders (e.g., sleep disorders), frequently coexist in elderly and chronic patients. Therefore, a growing demand for a more accessible healthcare system is arising from all healthcare stakeholders, both for improving people's quality of life and for reducing healthcare-related costs.

In order to meet these new needs, suitable technological solutions, generally referred to as e-health, have been studied. Traditional services include telesurgery and remote monitoring, which  are expected to significantly advance (with a more extensive use of virtual reality) in the near future. However, with the availability of pervasive mobile data communication networks, a new mobile health (m-health) paradigm is emerging, exploiting for example wearables for patient monitoring and virtual reality solutions for connected ambulances~\cite{5gpp,Soldani2017}. Indeed, the diffusion of m-health solutions should lead to the creation of a novel Internet of medical things (IoMT)~\cite{Irfan2018, Jusak2016}, with the promise of an ubiquitous, continuous, and personalized medical assistance, able of improving the living conditions of patients and reducing healthcare costs. 

However, m-health services may be life-critical, thus requiring high reliability and reduced latency. Early development of e-health services has highlighted the limitation of the  current mobile communication infrastructure, and new networks are needed. For example,  a tele-echography platform has been implemented over a long-term evolution (LTE) cellular link, providing rates around $10$~Mb/s, latency of $60$~ms and jitter of $8$~ms, still far from those provided by wired networks~\cite{Avgousti}.    

The new fifth generation (5G) generation of cellular networks promises to significantly improve over LTE, in particular about data rate, reliability, and latency. Indeed, since the earliest phases of 5G standardization in 2015, e- and m-health applications have been identified as relevant and challenging targets~\cite{5gpp}. Even more futuristic heathcare solutions are now being envisioned for the sixth generation (6G) of mobile networks~\cite{Saad-feb19}.

Although attention to m-health has been dedicated by some literature that sparsely  studied specific communication requirements, no in-depth analysis is available, differently from industrial automation applications~\cite{TS22261}. This paper aims at filling this gap, offering a comprehensive and structured overview of the most relevant health-related services, defining their technical requirements, finally achieving  a synthesis similar to that of~\cite{TS22261}. This will facilitate the definition and deployment of 5G networks. Moreover, we will also indicate further enhancements that will be made possible by 6G networks.

\section*{5G performance indicators}

We first describe the technical performance metrics used for standardizing the 5G cellular communication systems by the {\it Services and system aspects} group of the third generation partnership project (3GPP)~\cite{TS22261}. They have been already specified for automation and control services, e.g., discrete automation motion control, electricity distribution, and remote control. Then, we will use these metrics to characterize new healthcare services. The considered metrics are the following (see \cite{TS22261} for more details):

{\it End-to-end latency}:  average latency at application level. 5G targets an average latency down to $1$~ms, specifically intended for highly interactive applications, while in LTE the average latency is well above $20$~ms.

{\it Jitter}: measure of latency uncertainty, due to errors and non-idealities in communications. Jitter is quite significant in LTE networks, increasing end-to-end latency in many cases to more than $100$~ms.

{\it Survival time}: maximum packet delay still tolerated by the application to operate properly.

{\it Communication service availability}: percentage of the time in which the communication service is {\it up}. The service is considered in {\it down}-state when a packet is not received within a given time constraint (larger than the sum of end-to-end latency, jitter and survival time). 

{\it Reliability}: {\it degree of reliability}, i.e., the probability of successfully deliver network-layer packets to a given system entity within the time constraint required by the targeted service.

{\it User experienced data rate}: minimum acceptable data rate to satisfy the user experience of a specific service. In fact, 5G aims at improving over the LTE data rate by 10 to 100 times. 

{\it Payload size}: average size of the data packet. Addressing a wide variety of use cases, 5G has broaden the range of packet sizes. Here, we will refer to as {\it small} payloads when the packet is smaller than $256$~b, {\it medium} when it is between $256$~b and $10$~kb, and {\it big} when it exceeds $10$~kb.

{\it Traffic density}: communication data rate per unit area, based on the assumption that all applications within the area require a given user experienced data rate. 5G targets $1\,000$ times higher traffic density than LTE.

{\it Connection density}: number of connected devices per unit area, under the assumption of 5G full penetration.

{\it Service area dimension}: volume of the geographic area where the service is available. 

Besides, when dealing with healthcare, the following additional metric could be useful:

{\it Distance}: distance between connected entities, e.g., between the surgeon and the patient in a telesurgery system. The distance can significantly impact on other important parameters, such as latency.

\section*{Future healthcare services requirements} 

Among a broad range of potential healthcare services, we identified the most promising ones for their social impact and challenging for their communications requirements.
In particular, we grouped them into four major  scenarios, requiring specific technical communications features, either in terms of latency, massive access, mobility, or a combination of them. The scenarios are: a) telepresence and robotic telesurgery (including telesurgery and wireless service robots); b) remote pervasive monitoring (both in hospital and at home); c) healthcare for rural areas; and d) m-health (including the use of wearables and connected ambulance). 

It is worth noting that some of these scenarios could share key technologies   with scenarios in other fields. For example, tele-rehabilitation could make use of {\it virtual reality} (VR) technology to enhance the training effectiveness in patients who suffered from stroke, while VR  finds applications in other realms, from gaming to industry automation, too.
%

In the following, the four scenarios are described in more details, while Tables~\ref{tabella1}~and~\ref{tabella2} summarize their communication requirements, similarly to the tables prepared for other applications by 3GPP~\cite[Tables 7.1-1 and 7.2.2-1]{TS22261}. For each scenario we also indicate the relevant literature from which the requirements have been obtained. All these scenarios are connected with each others and can be viewed as a {\it medical network}.

\renewcommand{\arraystretch}{1.3}
\begin{table*}
\caption{Performance requirements for future healthcare applications -- Part I}
\label{tabella1}
{\centering
\small
\begin{tabular}{ | P{2.3cm} || P{2.5cm}| P{1.7cm} | P{1.0cm}| P{1.2cm} | P{1.7cm}| P{1.5cm} |  P{2.0cm} |}

\hline
\textbf{Scenario} & \textbf{Type} & \textbf{End-to-end latency} & \textbf{Jitter} & \textbf{Survival time} & \textbf{Commun. service availability} & \textbf{Reliability}   & \textbf{User experienced data rate}\\
&   & ms & ms & ms &    &     &  \\
\hline\hline

\multirow{2}{*}[-0.5em]{\parbox{2.3cm}{\centering 
Telepresence and robotic\\ telesurgery}}  & Telesurgery & 5 & 2 & \multirow{2}{*}[-0.7em]{0} 
				& \multirow{2}{*}[-0.7em]{$1-10^{-5}$} & \multirow{2}{*}[-0.7em]{$1- 10^{-7}$} & $2$~Gb/s \\   
\cline{2-4,8} 
					& Wireless service robots		& 1 & 1 &   &   &  & $1$~Gb/s \\ 
\hline

\multirow{2}{*}{\parbox{2.3cm}{\centering 
Remote pervasive monitoring \\~}}  & In hospital & \multirow{2}{*}{250} & \multirow{2}{*}{25} & \multirow{2}{*}{10} & \multirow{2}{*}{$1-10^{-3}$} & \multirow{2}{*}{$1- 10^{-3}$} & $300$~Mb/s \\
\cline{2,8} 
					& At home		&  &  &   &   &  & $1$~Gb/s \\ 
\hline

Healthcare in rural areas & - & 20 & 10 &  10 & $1-10^{-3}$ & $1-10^{-3}$ & $1$~Gb/s \\
\hline

\multirow{2}{*}[-0.7em]{
m-health$^1$} & Wearables 			& 250 & 25 & 10 & $1-10^{-3}$ & $1-10^{-3}$ & $0.1\div 5$~Mb/s  \\
\cline{2-6,7-8}
    		& Connected ambulance  			& 10 & 2 & $<2$ & $1-10^{-5}$ & $1-10^{-7}$  & $50 \div 1\,000$~Mb/s \\
\hline
\end{tabular}
}
~\\[0.05cm]
$^1$ Performance requirements must be ensured for speeds up to $50$~km/h (wearables) or $100$~km/h (ambulance).
\end{table*}

\subsection*{Telesurgery}

Computer-aided surgery and, particularly, {\it telesurgery}, is seeing a rapid development and a spreading of use, given its many advantages over traditional surgery. Indeed, it enables 1) visual guidance during surgery, 2) augmented reality (AR), helpful for the surgery team, and 3) the assistance by several robots that highly reduce invasiveness, thus making the patient recovery faster.

In order to make telesurgery remotely available, highly demanding communications requirements have been recently identified~\cite{Soldani2017, Zhang2018}. 
Indeed, in order to experience a natural perception during surgery, physicians should be supported by a stable, real-time, haptic, and visual feedback coming from the surgical device and the camera pointing at the anatomical target: this requires data rates of at least 1~Gb/s~\cite{Zhang2018}, medium to short latencies (from $200$~ms to less than $5$~ms)~\cite{5gpp,Zhang2018}), small jitters (from $30$~ms to less than $2$~ms)~\cite{Zhang2018}, and the reliability should not be lower than $(1-10^{-7})$~\cite{Soldani2017}, with low traffic connection density. Distance may reach 300~km, making the latency requirement quite challenging.

Furthermore, in some recent attempts, the integration of holographic models of the target anatomical structure has been proposed. In this case, requirements become further demanding: indeed, the real-time holographic rendering needs end-to-end latencies below 5~ms~\cite{Zhao2016, Soldani2017}. At the same time, the user data rate could easily jump to 2~Gb/s.

\subsection*{Wireless Service Robots}
Telepresence defines a broad range of robotic services, where {\it wireless service robots (WSRs)} have a key role in providing assistance, support, monitoring, and even company to patients and elderly, in a large variety of indoor and outdoor environments. Including wellness into healthcare, WSRs are also increasingly being employed in advanced sport training setups. In this respect, seamless assistance is a key aspect, which could facilitate user compliance, especially in elderly, and enable large-scale acceptance in the future.
 
In order to provide seamless and natural interaction with users, massive collections of images, videos, audio tracks and environmental and motion sensor data have to be processed in real-time (often, in the {\it cloud}) and returned to the WSR to perform actions. 
In this scenario, data rates of the order of $1$~Gb/s are needed, together with latency of about $1$~ms~\cite{Simsek2016}, and a reliability of $(1-10^{-7})$~\cite{Soldani2017}.

WSRs are expected to spread in the near future, thus offering a high traffic density at medium connection density. 5G should enable machine-type massive access to the network for robots to communicate with clinical centers and other robots within a range of up to $50$~km.

\begin{table*}
\caption{Performance requirements for future healthcare applications -- Part II}
\label{tabella2}
{\centering
\small
\begin{tabular}{ | P{2.3cm} || P{2.5cm}| P{1.3cm} | P{1.5cm}| P{1.8cm} | P{2.1cm}| P{1.3cm} | P{1.8cm} | } 

\hline
\textbf{Scenario} & \textbf{Type} & \textbf{Payload size} & \textbf{Traffic density}  & \textbf{Connection density} & \textbf{Service area dimension} & \textbf{Distance} & \textbf{Reference} \\
 &   &   & Gb/s/km$^2$ & devices/km$^2$ & m $\times$ m $\times$ m &  km &\\
\hline\hline


\multirow{2}{*}[-0.5em]{\parbox{2.3cm}{\centering 
Telepresence and robotic \\telesurgery}}  & Telesurgery & big & low & low & $10\times 10\times 5$ & 300 & \cite{5gpp, Soldani2017, Zhang2018, Zhao2016} \\ 
\cline{2-8} 
			& Wireless service robots		& medium &  high &  medium & $100\times 100\times 3$  & 50  &  \cite{Soldani2017,Simsek2016} \\ 
\hline

\multirow{2}{*}[-0.5em]{\parbox{2.3cm}{\centering 
Remote pervasive monitoring}}  & In hospital & medium & $1$ & $1\,000$ & $10^3 \times 10^3 \times 50$ & $1$ & \cite{Thuemmler2016, Zhang2018} \\
\cline{2-8}
			& At home		& small & $< 10^{-3}$  & $20\,000$  & $10^3 \times 10^3 \times 50$  & 20 & \cite{Jusak2016, Alfian2018, Malasinghe2019, Zhang2018} \\ 
\hline

Healthcare in rural areas & - & medium & low & low & $ 10^4 \times 10^4 \times 10$ & $100$ & \cite{5gpp,Islam2015} \\
\hline

\multirow{2}{*}[-0.7em]{m-health$^1$} & Wearables & small  & 50 & $10\,000$ & $1 \times 1  \times 2$ & 50 & \cite{Zhang2018,5gpp,Islam2015,Malasinghe2019} \\ 
\cline{2-8}
& Connected ambulance & big & low & low & $5 \times 1.5 \times 2.5$ & $50$ &\cite{5gtrials} \\
\hline

\end{tabular}
}
~\\[0.05cm]
$^1$ Performance requirements must be ensured for speeds up to 50~km/h (wearables) or 100~km/h (ambulance).\\
\end{table*}

\subsection*{Remote Pervasive Monitoring in Hospital}
In this scenario, heterogeneous, yet complex, continuous, and ubiquitous (thus pervasive) monitoring is provided to many inpatients. A large amount of data is collected from both routine (e.g., Holter's electro-cardiogram, ECG) and sophisticated devices (e.g., magnetic resonance system). Data are integrated, in real-time, to determine the current health condition of the patient, and in critical situations, an assistance request can be automatically triggered. The patient health-track history is also considered, in order to reduce the false-alarm probability, while maintaining a comparable mis-detection probability. Various hospital units could benefit from this technology. For example, the intensive care unit could continuously monitor patients' blood pressure, oxygenation, heart beat rate, and other vitals, to predict possible heart attacks. 
Pervasive monitoring generates traffic with data rates up to $1$~Mb/s/patient~\cite{Zhang2018}.
In general units, physiological data of patients may be continuously collected and processed (in the {\it cloud}) to maintain a reliable picture of the patient's status.
Therefore, the high connection and traffic density given by a high number of monitored patients (expected to be $100$ per unit in 2035,~\cite{Thuemmler2016}) within the limited area of the hospital can challenge the communication network. The overall data rate could stably approach $300$~Mb/s (including imagery data~\cite{Thuemmler2016}), while moderate requirements on latency ($250$~ms), jitter ($25$~ms), survival time ($10$~ms), and service availability ($1-10^{-3}$) are expected.

\subsection*{Remote Pervasive Monitoring at Home}

Pervasive monitoring can be extended from hospital to the individual's home~\cite{Alfian2018,Malasinghe2019}. In this scenario, cheaper, yet reliable, biosensors can be provided to the user to monitor for the vitals and other physiological activities, e.g., the glucose level~\cite{Zhang2018}.
Moreover, a {\it smart} house can be equipped with additional tiny and unobtrusive sensors for measuring temperature, humidity, and light exposure data. In some cases, cameras can monitor the patient activity, especially for those at risk of falling.
Biosensors can include traditional acquisition systems, such as Holter's ECG, but also wearables and portable devices that can be organized in an individual's wireless body area sensor network (WBASN), which collects and periodically sends patients' data to a hospital, a clinical center, or even an ambulance, for both real-time monitoring and data storage.
Pervasive monitoring at home includes teleconsultations and other speech/cognitive/physical therapy sessions (assisted by a remote clinician), reducing the burden of regularly reaching the hospital for therapy. 
%
%
%

Here, real-time continuous and heterogeneous big data collection, transmission, analysis, and feedback must be performed by the medical network, in fractions of seconds, in order to generate alarms when needed, update the diagnosis, and provide indications to improve the training.
Therefore, the latency requirement, together with the massive data transmission, could challenge the network performance: easily, 1 Gb/s traffic can be generated by monitoring a densely populated urban area (e.g., with over $2 000$~inhabitants/km$^2$)~\cite{Jusak2016}.
Moreover, with 5G, a specific priority could be accommodated for different pervasive monitoring at-home modalities, e.g., video streaming or alarms, by using ad-hoc {\it network slices}, offering different quality of service (QoS) levels.
%

\subsection*{Healthcare in Rural Areas}
Sparse {\it points-of-care} are intended to serve patients in rural, remote or underdeveloped areas, where specialized physicians are hardly present (thus with low traffic and connection density). At those sites, general physicians may be available for patients basic examination with simple equipment. Nevertheless, they could receive teleconsultation by more expert colleagues through videoconferencing systems; in addition, advanced examinations could be conducted by the remote expert clinician with the help of the local physician.

In this scenario, an advanced wireless communication network is needed to provide seamless remote video interaction with both the local physician and the patient, and to collect data from biosensors and other examination equipment through, e.g., body images~\cite{5gpp,Islam2015}.
Considering a rural area with a population density of few tens of inhabitants/km$^2$, the traffic could require user experienced data rates between $100$~Mb/s (in absence of video consultation) and $1$~Gb/s (when video streaming is needed).
Acceptable latencies can range between $150$~ms and $20$~ms (when real-time multimedia stream is needed). In this scenario, the communication distance could be as long as $100$~km, leading to a challenging network design to ensure also a reliability of at least $(1-10^{-3})$.

\subsection*{Wearables}
Wearables are recently seeing a tremendous success, spreading not only for sport training and fitness purposes, but also in healthcare, thanks to their user-friendly interfaces, their unobtrusiveness and, even, the current trend.
Among others, {\it smart clothing} and accessories (e.g., watches and bracelets)  equipped with several tens of micro-biosensors can measure both vitals and physiological signals~\cite{Islam2015,Malasinghe2019}, offering a promising m-health scenario.

Wearables will generate a huge data traffic through the network, also to the {\it cloud}, where engines based on artificial intelligence (AI) could perform smart computing, provide suggestions to improve training and to ameliorate the individual's behaviour (e.g., sleep quality). As healthcare solutions, implants are also considered as wearable devices. Their functionality can be controlled and modified by remote: for example, diabetic patients could have their glucose level regularly monitored~\cite{Zhang2018,5gpp}. 
%

Multiple wearables could be handled by a WBASN, where a smartphone could serve as router and primary collector for many and heterogeneous data.
In this scenario, the communication network must provide a stable and reliable link while an (expected) high number of users is freely moving in a variety environments (e.g., indoor as well as outdoor), at different speeds (e.g., walking, cycling, or even driving the car).
Heterogeneity in data traffic together with  high mobility (up to $100$~km/h), within distances up to $50$~km, are among the most important challenges  from a communication perspective.

\subsection*{Connected Ambulance}
In recent 5G trials, a {\it connected ambulance} has been equipped and tested on a 4G-enhanced network~\cite{5gtrials}. The equipment included a 4K camera providing detailed real-time videos of the patient with high color resolution (critically important for a reliable diagnosis) to clinicians assisting the paramedical staff from the hospital.
These solutions, as teleconference, require high data rates and low-latency, in order to  guarantee smooth streaming at moderately high speeds (up to $100$~km/h).
Preliminary clinical examinations, e.g., ECG, can be conducted on-board and by real-time teleconsultation, physicians from the hospital can provide urgent indications to paramedical staff in the ambulance.  At the same time, paramedicals could also perform simple maneuvers on patients under the step-by-step guidance of an autonomous intelligent system (running in the {\it cloud}) on the basis of monitored vitals and physiological indicators. As an example, the system could acquire data from the patient through smart glasses worn by the paramedicals and return them using an AR feedback that integrates the vision of the patients as well as the above mentioned instructions. Low latency and jitter are needed for a proper feedback.
In this scenario, a very small survival time has to be ensured (e.g., $2$~ms), a high connection reliability, despite the ambulance is running at high speeds over  distances up to $50$~km.

\section*{Matching 5G to Healthcare Needs}
The international telecommunication union (ITU) defined three general target categories for 5G cellular networks: the enhanced mobile broadband (eMBB), providing high data rate to mobile human users, the ultra-reliable-low-latency communications (URLLC), typically targeting human-machine interactions, and the massive machine-type communications (mMTC), for autonomous interaction among devices.
In healthcare, as previously discussed, stringent requirements are mainly related to latency, user experienced data rate, massive access, and mobility. Nevertheless, the scenarios described above fall at the intersection between the three 5G target categories. For example, healthcare-related mobility in outdoor environments comes across URLLC and mMTC: e.g., a portable insulin-pump closed-loop system for diabetic patients needs  an ultra-reliable network (URLLC) for communications between devices (mMTC) that continuously and autonomously adjust the insulin injection based on the patient's status.

\begin{figure*}
	\centering
	\includegraphics[width=0.98\textwidth]{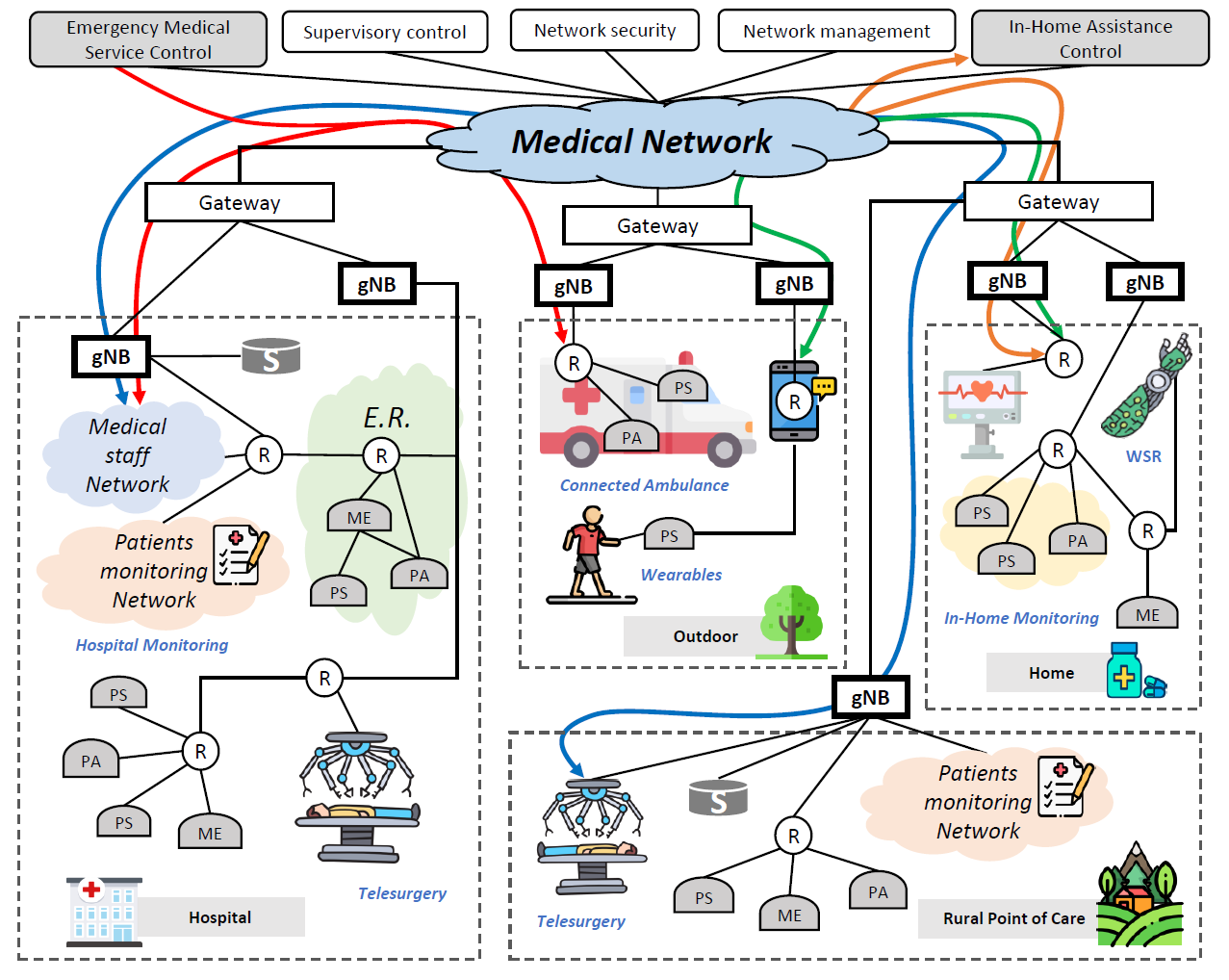}
	\caption{Network scenario and communication paths for service flows of the health-care network. R: router; S: server; ME: medical equipment; PS: patient sensor; PA: patient actuator.} \label{fig:IoMTnetwork}
\end{figure*}

A summary of the performance requirements for the various scenarios is provided by Tables~\ref{tabella1}~and~\ref{tabella2}:  it becomes clear that 5G systems will match the requirements for the various healthcare scenarios.
A pictorial description of the scenarios is provided by Fig.~\ref{fig:IoMTnetwork}, similarly to   Fig.s D1-1 and D.2-1 of~\cite{TS22261} relative to other 5G use cases.
Here, the described scenarios are organized into four physical spaces: a rich and technologically advanced environment (e.g., a hospital), an indoor space, e.g., a smart home, an outdoor area, and a rural location.
The medical network, possibly exploiting {\it cloud} computing, is expected to connect all healthcare-related services. It includes control entities for specific healthcare-related tasks (e.g., emergency management and home assistance), but also communication entities for network supervision, security, and management.
Connection among devices is provided by the 5G cellular network through data gateways and new-generation node-bases (gNBs).
End devices may also be connected to the network via routers (R), creating WBASNs in smart homes,  rural-area point-of-cares, and dedicated emergency rooms (ERs) connected to ambulances. Devices include WSRs, generic medical equipments (MEs), patient sensors (PSs), patient actuators (PAs) for delivering drugs and regular medications, and telesurgery equipment. Servers (S) are available at local networks to store patient data and software needed for medical network functioning and edge-computing of big data, including AI functions.
Various communication paths can be identified in the network: red lines represent communication flow associated to an emergency event managed by the emergency medical service control, operated via a connected ambulance, and solved by the medical staff at the hospital.
The blue line refers to the communication flow of a telesurgery with the patient in a rural area and the specialized surgeon at a remote hospital.
The orange line shows the communication path for monitoring patients at their smart home: the service is managed by the in-home assistance control unit and the end devices (i.e., biosensors and WSRs) collect data and send it, through the medical network, to the control unit providing intervention if needed.
Lastly, the green line connects the outdoor devices (possibly organized as a WBASN) to the smart home network through a smartphone, operating as a router, and the medical network, where {\it cloud}-computing is available.

From the figure it is clear how the inter-play among different network entities and healthcare services   composes a truly complex picture that needs to be carefully and efficiently handled with the support of  the future 5G cellular networks.
%
%
One of the most challenging and common characteristic of the health-related data traffic is its heterogeneity. Extremely low-latency and high-data-rate telesurgery video traffic could coexist with much lower-data-rate, but massive, traffic by plenty of wearables and portable devices in the same location, e.g., a technologically advanced hospital.
Therefore, 5G will significantly advance LTE by accommodating differentiated traffics, aggregating data, and  automatically setting priorities, by either edge-AI-based computing and network slicing. This will spur new heterogeneous healthcare services, pushing forward the development of a new paradigm of personalized and ubiquitous medicine, currently strongly targeted by the most healthcare stakeholders. 
 
\subsection*{Beyond 5G Networks}

Although 5G will be offer a leap forward for the communication support of m- and e-health services, paving the way for a IoMT, new even more advanced and challenging features are envisioned for the future. In this sense, the immersive experience of VR, critical for both telesurgery and connected ambulances, can be further enhanced by eXtended reality (XR), including aumented, virtual, and mixed reality with $360^{\circ}$ high-quality vision. XR is seen as one of the key components of 6G systems~\cite{Saad-feb19}, which is still at their earliest stage of definition. Moreover, other 6G scenarios include the spreading of new sensors, wearables, implants, and wireless devices for brain-computer interfaces~\cite{Saad-feb19}, that would further raise the communications requirements (in particular, latency and connection density) of 6G networks.

\section*{Conclusions}
The urgency for new healthcare services to cope with the rapidly aging worldwide society and the related needs of an ever-increasing number of chronic patients would highly benefit from the development of new 5G cellular networks in the near future.
Indeed, 5G promises to strongly increase data rates and reduce both end-to-end latency and jitter, while ensuring high reliability. From our analysis we conclude that 5G will truly spur new e- and m-health applications, offering a personalized and ubiquitous medicine, while even more innovative scenarios are already at the horizon of 6G networks.

\section*{Biographies}

\small

{\bf Stefano Tomasin} (tomasin@dei.unipd.it) is with the Department of Information Engineering of the University of Padova, Italy. His current research interests include physical layer security and signal processing for wireless communications, with application to 5th generation cellular systems. In 2011-2017 he has been Editor of the IEEE Transactions of Vehicular Technologies and since 2016 he is Editor of IEEE Transactions on Signal Processing. Since 2011 he is also Editor of EURASIP Journal of Wireless Communications and Networking.

{\bf Giulia Cisotto} (cisottog@dei.unipd.it) received her Ph.D. in 2014. Currently, she is assistant professor (non-tenure track) at the Department of Information Engineering of the University of Padova (Italy). Since 2010, she had worked on EEG and EMG signal processing for neuro-motor rehabilitation, including brain-computer interface for stroke and dystonia patients. Her current research focuses on signal processing for electrophysiological signals, including machine learning, e-health technologies, and Internet-of-Medical-Things.
 
{\bf Edoardo Casarin} (edoardo.casarin@studenti.unipd.it) is currently pursuing his master degree in ICT for Internet and Multimedia at the Department of Information Engineering of the University of Padova, where he obtained in 2019 the bachelor degree in biomedical engineering. His studies focuses on {\it Life and Health} technologies.

\end{document}